\documentclass[aps,prl,twocolumn,superscriptaddress,showpacs]{revtex4}
\usepackage{graphicx}
\usepackage{color}
\usepackage{amsmath}
\usepackage{amssymb}

\begin{document}
\title{Expansion of a quantum gas released from an optical lattice}
\author{F. Gerbier}
\email{fabrice.gerbier@lkb.ens.fr}\affiliation{Laboratoire Kastler
Brossel, ENS, UPMC, CNRS ; 24 rue Lhomond, 75005 Paris, France}

\author{S. Trotzky}
\affiliation{Institut f{\"u}r Physik, Johannes
Gutenberg-Universit{\"a}t, 55099 Mainz, Germany.}

\author{S. F{\"o}lling}
\affiliation{Department of Physics, Harvard
University, Cambridge, MA 02138, USA.} 

\author{U. Schnorrberger}
\affiliation{Institut f{\"u}r Physik, Johannes
Gutenberg-Universit{\"a}t, 55099 Mainz, Germany.}

\author{J. D. Thompson}
\affiliation{Institut f{\"u}r Physik, Johannes
Gutenberg-Universit{\"a}t, 55099 Mainz, Germany.}

\author{A. Widera}
\affiliation{Institut f\"ur Angewandte Physik,
53115 Bonn, Germany} 

\author{I. Bloch}
\affiliation{Institut f{\"u}r Physik, Johannes
Gutenberg-Universit{\"a}t, 55099 Mainz, Germany.}

\author{L. Pollet}
\affiliation{Theoretische Physik, ETH Zurich, 8093 Zurich, Switzerland}

\author{M. Troyer}

\affiliation{Theoretische Physik, ETH Zurich, 8093 Zurich, Switzerland}

\author{B. Capogrosso-Sansone}
\affiliation{Department of Physics, University of Massachusetts, Amherst, MA
01003, USA}

\author{N. V. Prokof'ev}
\affiliation{Theoretische Physik, ETH Zurich, 8093 Zurich, Switzerland}\affiliation{Department of Physics, University of Massachusetts, Amherst, MA
01003, USA}
\affiliation{Russian Research Center ``Kurchatov Institute'', 123182 Moscow,
Russia}

\author{B. V. Svistunov}
\affiliation{Department of Physics, University of Massachusetts, Amherst, MA
01003, USA}
\affiliation{Russian Research Center ``Kurchatov Institute'', 123182 Moscow,
Russia}

\date{\today}
\begin{abstract}
We analyze the interference pattern produced by ultracold atoms
released from an optical lattice. Such interference patterns are commonly interpreted
as the momentum distributions of the trapped quantum gas. We show that
for finite time-of-flights the resulting density distribution can,
however, be significantly altered, similar to a near-field diffraction regime
in  optics. We illustrate our findings with a simple model and realistic 
quantum Monte Carlo simulations for bosonic atoms, and compare the latter to experiments.

\end{abstract}
 \maketitle
%
%
%

Experiments with ultracold quantum gases in optical lattices rely
heavily on time-of-flight (ToF) expansion to probe the spatial
coherence properties of the trapped gas
\cite{greiner2002a,stoeferle2004a,gerbier2005a,gerbier2005b,ospelkaus2006a,guenter2006a,spielman2007a,catani2008a}.
When the phase coherence length is large compared to the lattice
spacing, the post-expansion density distribution shows a sharp
interference pattern with the same symmetry as the reciprocal
lattice. As the phase coherence length decreases, {\it e.g.}, on
approaching the Mott insulator (MI) transition, the visibility of this
interference pattern decreases accordingly \cite{greiner2002a}. To
obtain a more precise understanding beyond this qualitative
description, it is usually assumed that the density distribution
$n_{\rm ToF}({\bf r})$ of freely expanding clouds provides a
faithful map of the initial momentum distribution.

In this Letter, we point out that, in general, the ToF distribution differs from the momentum distribution for finite time-of-flight,
the latter being recovered only in the ''far-field'' limit $t\rightarrow
\infty$. Practically, the ToF and momentum distributions become
identical after a characteristic expansion time $t_{\rm FF}= m R_0
l_c/\hbar$, which depends on the particle mass $m$, the coherence 
length $l_c$, and the cloud size $R_0$ prior to expansion. This time scale can be
understood in analogy with the diffraction of a coherent optical
wave by a periodic grating. Then, the characteristic $t_{\rm FF}$ in
the expansion problem exactly corresponds to the Fresnel distance in
the diffraction problem. The far-field regime is typically reached
when the coherence length is short, for example for a cloud in the MI regime, 
or a thermal gas well above the critical temperature. We show that for phase-coherent
samples where a sizeable fraction of the atoms are Bose condensed,
the far-field condition is usually not met for typical expansion
times used in current experiments
\cite{greiner2002a,stoeferle2004a,gerbier2005a,gerbier2005b,ospelkaus2006a,guenter2006a,spielman2007a,catani2008a}. Experimental measurements and quantum Monte-Carlo simulations are used to demonstrate that this results in substantial changes in the ToF distribution. We also discuss implications for the interpretation of the ToF images.

We consider an ultracold boson cloud released from a periodic trapping potential with cubic symmetry, lattice spacing $d=\lambda_L/2$, and lattice depth $V_0$ given in units of the
single-photon recoil energy $E_{\rm R}=h^2/2m\lambda_{\rm L}^2$,
where $\lambda_{\rm L}$ is the lattice laser wavelength. 
In addition to the lattice potential, an ``external'' harmonic potential is present, due to both the magnetic trap and the optical confinement provided by the Gaussian-shaped lattice beams \cite{gerbier2005b,greinerphd}. This external potential is responsible for the appearance of a shell structure of alternating MI and superfluid regions in the strongly interacting regime.

\begin{figure}[ht]
\includegraphics[width=8cm]{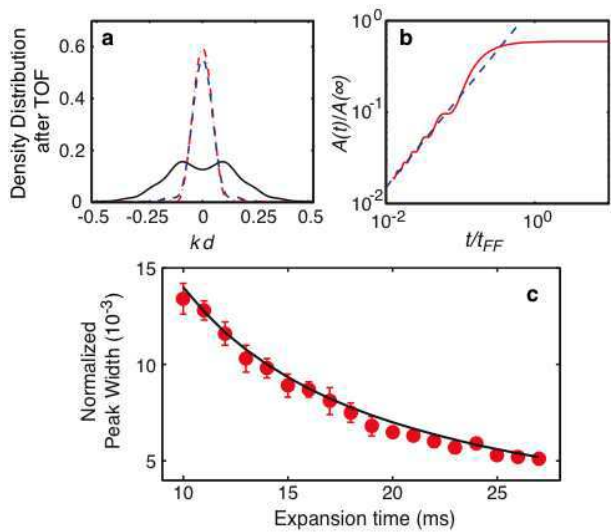}

\caption{{\bf (a)} Momentum distributions for a one-dimensional
lattice with parabolic distribution of the occupation numbers
calculated using Eq.~(\ref{S1d}) (solid line,
expansion time $t=20~$ms, dashed line: expansion time $t=100~$ms, dot-dashed line: expansion time $t \rightarrow \infty$).
{\bf (b)} Evolution of the peak amplitude $A$ with expansion time
$t/t_{\rm FF}$. The dashed line shows the expected near-field scaling in one dimension, $ A \propto t/t_{\rm
FF}$. The number of sites is $2N_{\rm TF}+1=61$ for (a) and (b). {\bf (c)}
Evolution of the width of the diffraction peaks with expansion time. The width has been normalized to the separation
between two adjacent diffraction peaks for convenience. The  circles show the experimental measurements and the solid line a fit by a hyperbola $\propto 1/t$,  as expected in the near-field.}
\label{vis2}\end{figure}

The density distribution after expansion for a time $t$ is usually expressed as a product 
(see, {\it e.g.}, \cite{pedri2001a}),
\begin{equation}\label{ntof}
n_{\rm ToF}({\bf r})=\left( \frac{m}{\hbar t} \right)^3 \left|
\tilde{w}_0({\bf k})\right|^2
\mathcal{S}\left({\bf k}\right),\;\;\text{with}\;\;{\bf k}=\frac{m {\bf r}}{\hbar t},
\end{equation}
where an envelope function $\tilde{w}_0$ is the Fourier transform of the
on-site Wannier function $w_0$) and the interference term is
\begin{equation}\label{Sk}
\mathcal{S}({\bf k})= \sum_{r_{\bf \mu},r_{\bf \nu}} e^{i{\bf
k}\cdot({\bf r}_\mu-{\bf r}_\nu)}\langle \hat{a}_\mu^\dagger
\hat{a}_\nu\rangle.
\end{equation}
Here the operator $\hat{a}_\mu^\dagger$ creates an atom at site
${\bf r}_\mu$. To assess the validity of the far-field approximation used in
Eq.~(\ref{ntof}), we quickly outline its derivation. Neglecting
interactions during expansion (see below), the atomic field operator
can be expressed in Schr\"odinger's picture as $
\hat{\Psi}({\bf r},t)=\sum_{r_{\bf \nu}} W_\nu({\bf r},t)
\hat{a}_\nu$ where $W_\nu({\bf r},t=0)=w_0({\bf r}-{\bf r}_\nu)$. After
the cloud is released, the wavefunction $W_\nu$ evolves in
free flight as $W_\nu({\bf r},t) \approx \left( \frac{m}{\hbar t}\right)^{3/2}
\tilde{w}_0\left(\frac{m ({\bf r}-{\bf r}_\nu)}{\hbar
t}\right)e^{i\frac{m ({\bf r}-{\bf r}_\nu)^2}{2 \hbar t}}$ for $\omega_L t \gg 1$, with  $\omega_L$ the oscillation frequency at the bottom of a lattice well.
In the limit $t \rightarrow \infty$, the dependence on the initial site position ${\bf r}_\nu$
vanishes, and one recovers Eq.~(\ref{ntof}). For finite
$t$, this dependence can be neglected in the envelope function \cite{footnoteTOF}, but not in the phase factor. We thus obtain a generalized interference term
\begin{equation}\label{Sk2}
S_t({\bf k})= \sum_{r_{\bf \mu},r_{\bf \nu}} e^{i{\bf k}\cdot({\bf
r}_\mu-{\bf r}_\nu)-i \frac{m}{2\hbar t}({\bf r}_\mu^2-{\bf
r}_\nu^2)}\langle \hat{a}_\mu^\dagger \hat{a}_\nu\rangle.
\end{equation}
Note that experimentally one observes a column distribution integrated along the probe direction, $S_\perp({\bf k}_\perp)=\int dk_z~|\tilde{w}(k_z)|^2 S_t({\bf
k})$. This is included in latter comparison with experiments, but in the following we base our discussion on Eq.~(\ref{Sk2}) for simplicity.

A fruitful analogy can be made with the theory of optical
diffraction. The formation of the interference
pattern results from the interference of many spherical matter waves
emitted from each lattice site, with phase relationships reflecting
the initial quantum state of the boson gas. We can exploit this
analogy further by defining the equivalent of a Fresnel distance
usually introduced in the theory of optical diffraction to estimate the importance of the quadratic phase factor $\propto {\bf r}_\mu^2-{\bf r}_\nu^2$. Because the correlation function $\langle
\hat{a}_\mu^\dagger \hat{a}_\nu\rangle$ suppresses contributions
from sites distant by more than the characteristic coherence length
$l_c$, we can estimate the magnitude of the quadratic phase in Eq.~(\ref{Sk2}) as $\frac{m}{2\hbar
t}({\bf r}_\mu^2-{\bf r}_\nu^2) \sim \frac{m l_c^2}{2\hbar t}$ near
the cloud center, and $\sim
\frac{m l_c R_0
 }{\hbar t}$  near the cloud edge. Here $R_0$ the characteristic size of the cloud before expansion. The most restrictive condition to apply the far-field
approximation thus reads $t\gg t_{\rm FF}$, with
\begin{eqnarray}
t_{\rm FF}\approx \frac{m l_c R_0}{\hbar}.
\end{eqnarray}
As an example, for a $^{87}$Rb condensate with $l_c \approx R_0
\approx 30~d$ and a lattice spacing $d\approx400~$nm, one finds
$t_{\rm FF}\approx 100~$ms, much larger than typical expansion times $t\approx 20~$ ms in experiments \cite{footnoteToth}. In contrast, a gas with short coherence length ({\emph e.g.}, in the MI regime), with $l_c
\gtrsim d$, will enter the far-field regime after a few ms. We stress that the quadratic Fresnel term is intrinsically non-local,  as the dephasing between two particular points ${\bf r}_\mu$ and ${\bf r}_\nu$ depends not only on their relative separation but also on their absolute positions. Although this has little effect deep in the superfluid or in the MI phase, this casts serious doubts on the validity of a local density approximation to compute quantitatively the ToF distribution in regimes where the coherence length is intermediate between the cloud radius and the lattice spacing.

To illustrate the influence of Fresnel terms on the interference pattern, we consider a 1D lattice with uniform
phase and parabolic distribution of the occupation numbers, $\langle
\hat{a}_\mu^\dagger\hat{a}_\nu\rangle=c_\mu c_\nu$ with
$c_\mu=\sqrt{1-\left(\mu/N_{\rm TF}\right)^2}$. The ToF distribution
is given by
\begin{eqnarray}\label{S1d}
S_{\rm t}(\tilde{k})=\frac{1}{(2N_{\rm
TF}+1)^2}\left|\sum_{l=-N_{\rm TF}}^{N_{\rm TF}} c_l e^{i \tilde{k}
l-i \beta^2 l^2/2} \right|^2,
\end{eqnarray}
with $\tilde{k}=k d$, $N_{\rm TF}=R_0/d=30$ the Thomas-Fermi
condensate size in lattice units, and where $\beta=\sqrt{md^2/\hbar
t}$. The normalization factor $(2 N_{\rm TF}+1)^2$ would give the peak amplitude if
the filling factor were uniform. We plot in Fig.~\ref{vis2}a the distributions corresponding to $t=20~$ms$\sim 0.2t_{\rm FF}$ which shows a significant broadening of the distribution for short time of flight when compared to the asymptotic result. For longer expansion times $t\sim t_{\rm FF}\sim 100~$ms, the far-field approximation is recovered to a good approximation.

Qualitatively, we expect from dimensional arguments that the peak width scales as $ (\beta N_{\rm TF})^2=t_{\rm FF}/t$ in the near-field, while approaching a constant value in the far-field. The peak height thus increases as $(t/t_{\rm FF})^\mathcal{D}$ in $\mathcal{D}$ dimensions. This is confirmed by the one-dimensional calculation shown in Fig.~\ref{vis2}b. This dependence provides a mean to check  the importance of near-field effects experimentally. For the measurement, a sample of roughly $10^5$ $^{87}$Rb  atoms has been been prepared in a
three-dimensional optical lattice with a depth $V_y=6~E_{\rm R}$,
and subsequently released for expansion\cite{footnoteLattice}.  After recording a series of absorption images for different expansion times, the width of the interference peaks was extracted using a Gaussian fit to the images. We plot
the results in Fig.~\ref{vis2}c, normalized to the separation between two diffraction peaks for convenience. The data confirms the $t_{\rm FF}/t$ scaling, indicating that the far-field asymptote is not reached even after the longest expansion
time available in the experiment.

We now discuss briefly the effect of interactions on the expansion,
and show that this is negligible compared to the finite ToF effect.
When the cloud has just been released from the lattice potential, each on-site wavefunction $W_\mu$ expands independently with a characteristic expansion time $\omega_L^{-1}$, until $t\approx t^\ast = \sqrt{\hbar/(\omega_L E_R)}$ where the wavefunctions expanding from neighboring sites start to overlap. At this time,  in the usual situation where $\omega_L t^\ast \gg 1$, the local density has dropped dramatically by a factor $(\omega_L t)^{-3}\ll 1$. Hence, the interaction energy converts into kinetic energy on
the time scale of a few oscillation periods only, and expansion becomes rapidly ballistic. The parameter controlling the importance of interactions is given by $\eta=\frac{U}{\hbar \omega_L}\approx \sqrt{8\pi}~ \frac{a_s
n_0}{\lambda_L} \left(\frac{V_0}{E_R}\right)^{1/4}$,
with $U$ being the on-site interaction energy. For typical parameters, $\eta$ is small (for instance $\eta\approx 0.05$ for $V_0=10~E_R$ and the experimental parameters of \cite{gerbier2005a}). Hence, we expect only small corrections to the non-interacting picture of ballistic expansion. This has been confirmed using a variational model of the expanding condensate wavefunction \cite{perezgarcia1996a}. This model predicts that the  ''Wannier'' envelope expands faster as compared to the non-interacting case, which does not affect the interference pattern, and picks up a site-dependent phase factor formally similar to the Fresnel term discussed previously, but with a very weak prefactor $\eta \ll 1$ which has negligible influence in practice. We conclude that interactions essentially contribute to the expansion of the on-site wavefunctions, without significant dephasing of the interference pattern.

\begin{figure}[ht]

\includegraphics[width=8cm]{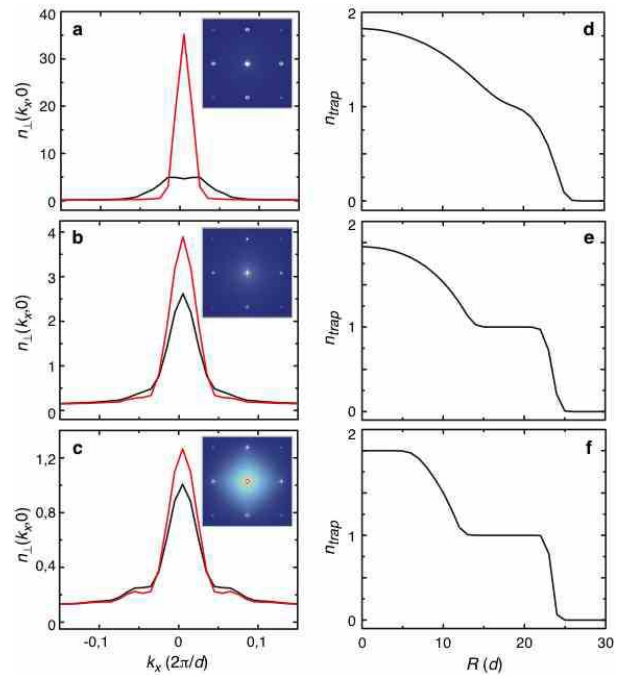}
\caption{Results from Quantum Monte Carlo simulations. On the left column, we show a horizontal cut through the ToF distributions for a finite expansion time $t=14~$ms (solid line), compared to a cut through the profile  calculated for $t \rightarrow \infty$ (dashed line). Units for $n_\perp$ are arbitrary. The insets show directly the two-dimensional ToF distributions for $t=14~$ms. On the right column, we show the in-trap density profiles for reference. The lattice depths are $V_0=12 E_R$ {\bf  (a,d)}, $15 E_R$ {\bf (b,e)} and $17 E_R$ {\bf (c,f)}, respectively. }
\label{profileQMC}\end{figure} 
The discussion so far focused on fully phase-coherent systems, which only applies to the weakly-interacting regime at low lattice depths. To investigate how the interference pattern is affected for strongly interacting systems ({\it i.e.,} on approaching the Mott transition and beyond), we have performed large-scale three-dimensional quantum Monte Carlo (QMC) simulations accounting for the external trapping potential using the worm algorithm~\cite{prokofev1998a,prokofev1998b} in the implementation of Ref.~\cite{pollet2007a}. The calculations were performed for $N=8 \times 10^4$ atoms, using exactly the same parameters and system sizes (up to $\sim 200^3$)  as in the experiments reported in \cite{gerbier2005a}. The simulation was done at low constant temperature $T = J/k_B$, where $J$ is the hopping amplitude. Although simulations at constant entropy would be closer to the experimental situation, the temperature turns out to be approximately constant in this parameter regime \cite{pollet2008a}.

The ToF distribution calculated for finite and infinite expansion times are shown in Fig. \ref{profileQMC}. The simulations confirm explicitly the analysis made above: the interference pattern is strongly affected in the superfluid phase, and the effect becomes less and less pronounced as the lattice depth is increased and the Mott transition crossed. Note finally that the Fresnel phase suppresses the contribution from the edges of the cloud, thus favoring the contribution of the central region to the ToF pattern. This is especially important when superfluid rings surround a central MI region with lower coherence \cite{batrouni2002a}.

\begin{figure}[ht]
 \includegraphics[width=8cm]{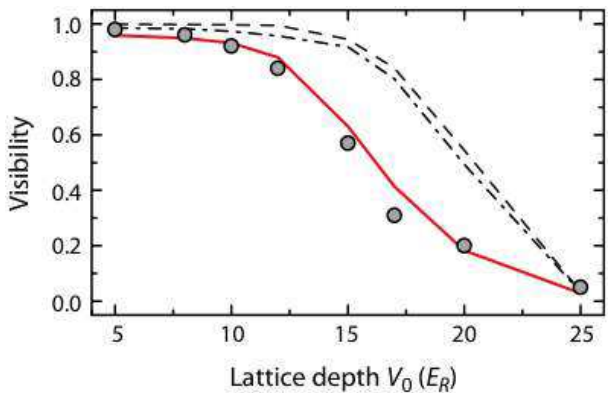}

\caption{Visibility of the interference pattern as defined in Eq. (\ref{V}). The dashed and dot-dashed lines show the Quantum Monte Carlo result for infinite and finite ($t=14$ ms) expansion times, assuming perfect experimental resolution. The solid line is computed for $t=14$ ms accounting for finite experimental resolution. Note that the comparison between experiment and simulation is only qualitative, since the simulations were performed at constant temperature $T=J/k_B$ while the experiment was not. }
\label{visibility}\end{figure}
The interference pattern is often characterized by its visibility \cite{gerbier2005a,gerbier2005b,ospelkaus2006a,guenter2006a,spielman2007a},
\begin{equation}\label{V}
\mathcal{V}=\frac{n_{\rm ToF}({\bf k}_{\rm max})-n_{\rm ToF}({\bf k}_{\rm min})}{n_{\rm ToF}({\bf k}_{\rm max})+n_{\rm ToF}({\bf k}_{\rm min})},
\end{equation}
with the choice ${\bf k}_{\rm max}d=(2\pi,0)$ and ${\bf k}_{\rm min}d=\sqrt{2}(\pi,\pi)$ to cancel out the Wannier envelope in the division. We first evaluate the sensitivity of $\mathcal{V}$ to the Fresnel phase by plotting in Fig.~(\ref{visibility}) two theoretical  ''benchmark'' curves assuming perfect experimental resolution (dashed and dot-dashed lines for $t=14~$ms and $t \rightarrow \infty$, respectively). We find little difference between the two curves
when $T/J$ is kept constant and small. Indeed, the Fresnel terms only matter for systems with large coherence length, where the visibility is by construction very close to unity.  We conclude that a detailed investigation of the superfluid side of the transition is better achieved by directly measuring the ToF distributions, whereas the visibility is well-suited for short coherence lengths.

We also compare in Fig.~\ref{visibility}  the experiments reported in \cite{gerbier2005b} to the predictions of the QMC simulations (solid line). Here, we emphasize that apart from the Fresnel terms, an accurate comparison requires to account for the experimental resolution, which is limited by two effects. First, the signal was obtained by integration over a square box centered around the maxima or minima, the integration area being $\approx (0.11\times 2\pi/d)^2$ in momentum units. This is comparable to a typical peak area, so that the visibility is calculated from the peak weight, rather than from its amplitude. Second, the finite resolution of the imaging system (about $6~\mu$m) is not negligible for the sharpest peaks. Accounting for these two effects when evaluating the QMC data, we find good agreement with the experimental results. This entails that the experimental data are compatible with the system remaining at low enough temperatures to cross a quantum-critical regime, in contrast to the analysis made in Refs.~\cite{diener2007a,kato2008a} which included neither near-field expansion nor experimental resolution.

In conclusion, we have analyzed the interference pattern observed
in the expansion of a bosonic quantum gas released from an
optical lattice. We showed that due to an additional Fresnel-like phase appearing for finite time of flight, the ToF distribution can be markedly different from the momentum distribution for clouds with large coherence lengths. Conversely, the visibility as calculated from Eq.~(\ref{V}) is rather insensitive to this effect. 

The Fresnel phase acts as a magnifying lens for the central region undergoing a Mott insulator  transition  by suppressing the contribution of the outer regions of the cloud when the central density is close to integer filling. This could eventually provide a way to investigate the physics near the quantum-critical point without ''parasitic'' contributions coming from coexisting superfluid rings.

Simulations were ran on the Brutus cluster at ETH Zurich. We acknowledge support from IFRAF, ANR (FG), DFG,  EU, AFOSR (IB), the Swiss National Science Foundation (LP),
NSF grant PHY-0653183 (BCS,NP,BS) and DARPA (OLE project).
 


\end{document}